\documentclass[sigconf]{acmart}
\AtBeginDocument{%
  }

\usepackage{enumerate}
\usepackage{graphicx}
\usepackage{float}

\setcopyright{acmlicensed}
\copyrightyear{2026}
\acmYear{2026}
\acmConference[...]{}{2026}{}
              
\acmISBN{978-1-4503-XXXX-X/2018/06}

\usepackage{amsmath}
\usepackage{listings}
\usepackage[dvipsnames]{xcolor}
\usepackage{tikz}
\usepackage{pgfplots}
\pgfplotsset{compat=1.18}

\usepackage{comment}
\excludecomment{omit}
\includecomment{comm}
\includecomment{blockcomm}
\specialcomment{blockcomm}{\noindent\bgroup\color{red}}{\egroup}

\newcommand{\OMIT}[1]{}
\newcommand{\mcomment}[2]{}
\newcommand{\footcomment}[2]{}
\newcommand{\margincomment}[2]{}
\newcommand{\ilcomm}[3][]{}

\newcommand{\wimb}[2][]{}
\newcommand{\ilpaolo}[2][]{}
\newcommand{\illeonid}[2][]{}
\newcommand{\ilmolham}[2][]{}

\newcounter{comment}

\begin{comm}
\renewcommand{\mcomment}[2]{\ifmmode\margincomment{#1}{#2}\else\footcomment{#1}{#2}\fi}
\renewcommand{\footcomment}[2]{{\color{blue}\textbf{(#1)}}\footnote{\textbf{#1:} #2}}
\renewcommand{\margincomment}[2]{{\color{blue}\textbf{(#1)}}\footnotemark\marginnote{\tiny\textsuperscript{\thefootnote}\textbf{#1:} #2}}
\renewcommand{\ilcomm}[3][blue]{{\color{#1}\textbf{#2}: #3}}
\renewcommand{\wimb}[2][blue]{\ilcomm[#1]{Wim}{#2}}
\renewcommand{\ilpaolo}[2][blue]{\ilcomm[#1]{Paolo}{#2}}
\renewcommand{\illeonid}[2][blue]{\ilcomm[#1]{Leonid}{#2}}
\renewcommand{\ilmolham}[2][blue]{\ilcomm[#1]{Molham}{#2}}
\end{comm}

\lstdefinelanguage{SQL}{
  morekeywords={
    SELECT,FROM,WHERE,JOIN,ON,AS,UNION,ALL,EXISTS,NOT
    WITH,RECURSIVE,INSERT,INTO,VALUES,UPDATE,SET,DISTINCT,EXCEPT,AND,PROJECT,IN,NOT,
    DELETE,CREATE,DROP,TABLE,COUNT,WITH,GROUP,BY,SUM
  },
  sensitive=false,
  morecomment=[l]{--},
  morestring=[b]'
}

\begin{document}

\title{Time to Move on:  Querying without Nulls and Bags}



\author{Molham Aref}
\affiliation{
\institution{RelationalAI}
\country{}
}

\author{Leonid Libkin}
\affiliation{%
 \institution{RelationalAI \\ University of Edinburgh}
 \country{}
 }

\author{Wim Martens}
\affiliation{%
 \institution{RelationalAI \\ University of Bayreuth}
 \country{}
}

\renewcommand{\shortauthors}{Aref, Libkin, Martens}

\begin{abstract}
SQL is \emph{the} database community's success story in terms of language design.
The key reason for its success 
is its declarativeness: it gives rise to optimizability, reducing the programmer’s burden significantly. 
However, given the evolving complexity of problems to solve
with query languages, our community needs to re-think some of the
fundamental early query language design decisions.

 Our experience in having worked on the design of Rel (a language for
 end-to-end relational programming) tells us that it is possible to
 design, implement, and successfully deploy a language based on fully
 normalized relations. Such relations avoid what Codd called
 \emph{corrupted relations} and what we commonly refer to as {\em bags},
 and the ``harmful'' \emph{billion dollar mistake} that we know as {\em
   nulls}. 
   In the SQL world, it is accepted that bags and nulls are
 tolerated as an unavoidable evil.  We argue
 that the evil is completely avoidable: reasons offered for justifying
 bags and nulls evaporate at a closer examination. In addition to
 debunking them, we also describe opportunities offered by a null-free
 language with set semantics. 

\end{abstract}

\keywords{Declarative languages, SQL, Nulls, Bags}


\maketitle

\section{Introduction}
SQL is the database community's greatest success story in terms of
language design. It goes back to 1973~\cite{Chamberlin12}, it became
an international standard in 1987, and remains extremely relevant
today. In addition to being a solid part of our database curricula, it
is often ranked as a top language required by
employers.\footnote{https://spectrum.ieee.org/top-programming-languages-2024
(or 2025), under "jobs".} The main reason for its success is arguably
its \emph{declarativity}: by disconnecting intent and execution, it
allows programmers to focus on getting the \emph{correct} results
instead of writing efficient programs to obtain them.


By now, SQL is a 40-year-old design, and as is normal for any design
so established, popular, and widely used, it has been dissected and
analyzed in excruciating detail. With many points of criticism and
known issues that programmers face daily, proposals for modifying and
improving SQL have also emerged.
In fact, early ideas of replacing SQL (or at least proposing an
alternative to it)  
are already found in Date and Darwen's Tutorial D \cite{DarwenDate1995}. This trend 
continued later with more modern proposals. Some of them, like 
\cite{prql,pipes}, try to sequentialize SQL and bring it closer to imperative programming; others, like Rel \cite{Rel}, SaneQL \cite{NeumannL-cidr24} fully adhere to the principle of declarativeness. 

A key criticism of SQL is the extremely high
complexity of SQL Standard. This complexity is inherent in its design as a {\em data sublanguage} \cite{Codd71}. That is to say, it is a declarative sublanguage for handling operations with data, with the rest delegated to a general purpose programming language. This approach made perfect
sense back in the 1970s and 1980s, 
when declarative programming had to win over alternative
proposals: 
going fully declarative at once was too ambitious. In fact, it is still an unrealized dream expressed by
Jim Gray as the holy grail of \emph{automatic
programming}~\cite{Gray03}, where the goal is to design a high-level
language sufficiently powerful for specifying \emph{all application
tasks}. 

SQL's sublanguage design, however, has led to problems. SQL lacks facilities for writing libraries, which entails that every new desired feature has to be added to the language
itself, and therefore become part of the standard. Today, the core of
the SQL Standard (SQL Foundation) is over 1,500 pages, and
the entire standard is well over 4,000 pages. For comparison, the
C standard is an order of magnitude smaller. A language that needs to be syntactically extended for each new
functionality ends up with hundreds of reserved keywords.

This makes queries error-prone: the more complex a language
is, the easier it is to make mistakes. Thus, some of the recent
language design efforts such as Rel~\cite{Rel} and SaneQL~\cite{NeumannL-cidr24} advocate creating a small
core of data operations. Most importantly, they both
preserve what is truly important about SQL: its declarative
nature. Beyond that, there are differences: Rel provides a small core
with the ability to write libraries and move beyond the data
sublanguage paradigm, while SaneQL keeps the data sublanguage approach
and fixes several glaring SQL issues to achieve a less error-prone language.

Where these languages differ most is in their approach to data
modeling. SQL relations are not the same as the
relations originally proposed by Codd. Those relations are {\em sets of
tuples of atomic values}. SQL violates this in two ways: first, its
relations are not tuples of atomic values, since they can contain
{\em nulls}, which require their own special logic; second, these relations
are not sets but {\em multisets (bags)}, so the same tuple can occur
multiple times. 

Rel eliminates both of these deviations and returns to Codd's original
model.  SaneQL, by contrast, treats them as a necessary
evil we have to live with. Our main goal is to show that this need not
be the case. This should not be seen as a criticism of SaneQL: we
think exploring different approaches to language design is an
extremely valuable exercise. Our goal is to argue that one need not
stick to solutions introduced by SQL four decades ago. 
Even if it is often claimed that we must accept nulls and bags as part of building a relational query language, most reasons offered why we need them 
can be debunked. This is precisely what we do here. 

We begin with a few quick examples showing how the logic of nulls and
bags differs from the logic programmers are used to, and how it can
easily lead to intuition-defying behavior. We then look at nulls and
three-valued logic, and debunk three myths: that nulls are needed to
model incomplete information; that they are needed to avoid
unnecessary joins when one chooses to eliminate nulls through full
normalization; and that three-valued logic is necessary to capture the
semantics of nulls. Having debunked these, we present the case for a
language with no nulls and no three-valued logic, showing that we can
dispense with both without losing expressiveness, while gaining more
natural queries, additional optimization opportunities, and more.

We then turn to bag semantics and examine the arguments offered in its
favor: that bags are necessary for correct aggregate computation, that
multiplicities may carry built-in semantics, and that bags are needed
for performance. We debunk these one by one. We then examine
how bags behave in more advanced languages needed for today's
workloads (in particular with recursive computation) and
show that bags interact disastrously with recursion, both in terms of
performance and semantics. We close the section by making the case
for sets, rather than bags, as the core data model. We show that
adopting sets uncovers additional optimizations lost under bag
semantics, and that it is necessary for the consistency of the
language and for a coherent understanding of what tuple counts
actually mean.
We also argue that simply
scattering {\tt DISTINCT} throughout SQL queries does not fix the problem.

\begin{figure}
\resizebox{.95\linewidth}{!}{
\begin{tikzpicture}
\begin{axis}[
    view={20}{30},
    shader=flat,
    axis lines=box,
    grid=major,
    domain=-3:3,
    y domain=-3:3,
    samples=40,
    samples y=40,
    xticklabels=\empty,
    yticklabels=\empty, 
    zticklabels=\empty
]
\addplot3[surf,
    draw=black!80,      
    fill=blue!30     
]
{
    0.65*exp(-(((x+1.2)^2 + (y+1.2)^2)/0.9))
  + 1.60*exp(-(((x-1.2)^2 + (y-1.2)^2)/0.9))
  + 0.10*exp(-((x^2 + y^2)/3))
};
\end{axis}
\coordinate (current) at (1.9,2.1);
\draw[fill=red!80!gray,draw=none] (current) circle (1mm);
\node (heretext) at (0.0,5.4) {We are here};
\draw[-,thick,red!80!gray]
    (heretext) to[bend right=20] (current);
    
\coordinate (localopt) at (2.1,2.42);
\draw[fill=red!80!gray,draw=none] (localopt) circle (.75mm);
\node (localopttext) at (4.1,4.8) {and are working hard to get here.};
\draw[-,thick,red!80!gray]
    (localopttext) to[bend left=20] (localopt);

\node[rotate=-5] at (2.2,.1) {\emph{some parameters}};
\node[rotate=45] at (6.3,0.8) {\emph{some parameters}};
\node at (7.6,4.8) {\emph{how good}};

\end{tikzpicture}
}
\caption{The Design Space of Query Languages}
\label{design-space:fig}
\end{figure}
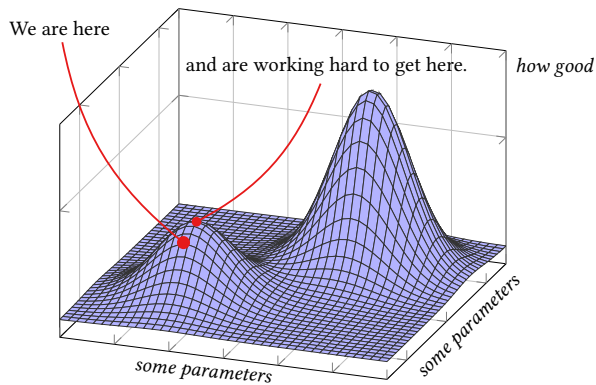

We end the introduction with the following thought. Consider
Fig.~\ref{design-space:fig}. It shows us where we are in terms of
query language design. With SQL and newer alternatives, we are not
that far off a local peak, but there are better options if we are
willing to leave our small local world. Instead of looking for
incremental improvements, we should consider climbing a different
hill: the one with a clean data model, no sublanguage paradigm, 
with
the ultimate goal of automatic programming at the top. The goal here
is to show that we can exit our local world and start a new climb by
dispensing with some heavy baggage that came in the shape of nulls and
bags.

\section{Losing Intuition With Bags and Nulls}

Multiple issues related to errors or intuition-defying behavior due to
duplicates or nulls have
been well-documented over the years. Here we present a couple, to set
the stage, before debunking some ``necessity myths''
about nulls and bags in the following sections.

\subsection{Intuition-Defying Nulls}
\label{bad-nulls:sec}

Consider relations \texttt{R(A)} and \texttt{S(A)} and three queries:
\begin{lstlisting}
SELECT DISTINCT A FROM R WHERE A NOT IN (SELECT * FROM S)

SELECT DISTINCT A FROM R 
WHERE NOT EXISTS (SELECT * FROM S WHERE S.A=R.A)

SELECT * FROM R EXCEPT SELECT * FROM S  
\end{lstlisting}
Assume that \texttt{R} and \texttt{S} contain no duplicates; all
queries use set semantics. Are they equivalent? Most people would say
yes, because they compute $\texttt{R} -
\texttt{S}$. This is true if no nulls are present, but with nulls, the
picture changes completely. If \verb+R={(1),(NULL)}+ and
\verb+S={(NULL)}+, the first query produces no tuples, the second outputs
the entire relation \verb|R|, and the third produces a single tuple
\verb|(1)|.
This happens due to two complementary ways of treating nulls. In the
first two queries, all comparisons involving nulls result in the truth value {\em  unknown}, and the condition in \texttt{WHERE} passes
only if it evaluates to {\em true}; thus in the first query nothing
gets selected while in the second, the subquery produces no tuples
making the \texttt{NOT EXISTS} condition true. On the other hand, in
set operations, nulls are treated {\em syntactically}: suddenly null
equals null, accounting for the result.




Nulls also invalidate some natural query equivalences that could be
useful for optimization. For example, $\sigma_{\theta}(R)\cup
\sigma_{\neg\theta}(R)$ need not equal $R$: the law of excluded middle
does not apply. Likewise, the \texttt{IN} predicate is not the same as
membership test: $a \in R$ is not captured by $a \ \texttt{IN}\  R$ in the
presence of nulls. Keeping the
three-valued logic,  varying treatment of nulls depending on query
operations, and their divergence from the common logic  in one's head is a tricky
proposition. It leads to many errors: ``You can never trust the answers you get from a
database with nulls'' \cite{Date2009SQL}.

\subsection{Intuition-Defying Bags}
\label{bad-bags:sec}

This is a classical example, but it never gets old. Suppose we have 
{\tt orders} and {\tt lineitem} relations, and we want to know how much
each customer spends on electronics items. One naturally writes this
query in SQL:

\begin{lstlisting}
SELECT o.cust_name, SUM(o.amount)
FROM orders o JOIN lineitem l ON o.oid=l.oid
WHERE l.category="electronics"
GROUP BY o.cust_name
\end{lstlisting}

Suppose that Jane placed one \$1000 order with two electronics items, a laptop and
a phone. The query output will then tell us that Jane spent \$2000,
because the join produces one row per matching lineitem, duplicating
the order amount before it gets summed. Such mistakes are easy to make
and hard to spot, as duplicates sneak in unexpectedly.

Similarly to nulls, bags also invalidate some query equivalences we take for granted over sets. For example, distributivity $R \cap (S \cup T) = (R\cap S) \cup (R \cap T)$ is no longer true if $\cap$ and $\cup$ are interpreted as {\tt INTERSECT ALL} and {\tt UNION ALL}.

\section{Nulls and 3-Valued Logic}

Nulls are considered "harmful"~\cite{Date24} and "fundamentally flawed"~\cite{Date2009SQL}. A related concept of null references in programming languages was branded a "billion dollar mistake" by Hoare, their inventor. So why do we use them? 
Arguments for nulls and three-valued logic (3VL) usually go along these lines:
\begin{enumerate}
\item We need nulls to model incomplete information. 
\item We need nulls to avoid joins.
\item 3VL is needed to capture the logic of nulls 
    and provide the necessary expressiveness to SQL queries.  
\end{enumerate}
We can debunk each of them.

\smallskip
\noindent \textbf{Debunk 1: Nulls are Needed for Incomplete Information.}
Many applications must deal with some sort of incomplete information. However,  SQL's \texttt{NULL} is hardly an ideal tool to handle it for at least two reasons. First, there are multiple types of incomplete information, and a single \texttt{NULL} is inadequate to represent them all.  Second, even if we agree that we only model one type of incomplete information, \texttt{NULL} can be completely avoided in the query language.

Concerning different types of incomplete information, a recent survey of database practitioners~\cite{ToussaintGLS22} revealed the following most common cases of the use of nulls (with 23 cases identified in total):
\begin{itemize}
\setlength{\itemindent}{-1em}
  \item \texttt{NULL} means an attribute is non-applicable.  
  \item \texttt{NULL} means we have no information whatsoever.  
  \item \texttt{NULL} means the value exists but is not known at the moment.  
  \item \texttt{NULL} means dirty data.  
  \item \texttt{NULL} means fixed known value that we do not want to record  
        (e.g.\ \texttt{0} to avoid runtime division errors, or confidential data). 
\end{itemize}
Thus, even if we put \texttt{NULL} in the database, it hardly tells us what is going on.  

Second, SQL's \texttt{NULL} can completely be avoided by fully normalizing the data into 6NF.
For example, the 6NF decomposition of 
\begin{center} 
{\tt 
\begin{tabular}{cc}
\toprule
A & B \\
\midrule
5   & NULL \\
NULL & 6   \\
7   & 8   \\
NULL & NULL \\
\bottomrule
\end{tabular}
}
\end{center}
has three relations with a \texttt{tuple\_id} attribute added:
\begin{center}
{\tt 
\begin{tabular}[t]{c}
\toprule
tuple\_id \\ \midrule id\_1 \\ id\_2 \\ id\_3 \\ id\_4 \\
\bottomrule
\end{tabular}
\qquad 
\begin{tabular}[t]{cc}
\toprule
tuple\_id & A \\
\midrule
id\_1 & 5 \\
id\_3 & 7 \\
\bottomrule
\end{tabular}
\qquad
\begin{tabular}[t]{ccc}
\toprule
tuple\_id & B \\
\midrule
id\_2 & 6 \\
id\_3 & 8 \\
\bottomrule
\end{tabular}
}
\end{center}
We can obtain the original table as the outer join of these.  This may be countered by a performance argument, which we address next.

\noindent \textbf{Debunk 2: We Need Nulls to Avoid Extra Joins.}
If we design a query language around fully normalized data as we just suggested, does not this lead to many extra (expensive) joins? We do not think so, and offer several orthogonal reasons. 

First and foremost, one major strength of query languages is \emph{data independence}: the user does not need to worry about how the data is stored.  In particular, even if the query language would treat relations as fully normalized, the engine does not necessarily need to. It can choose to store a relation however it wants, one option being a table with nulls. Notice that we are not contradicting ourselves here: we are saying that (a saner version of) nulls may have its place in \emph{lower-level languages for databases} (using outer joins, for example) and in {\em data structures}, but \emph{not in the highest-level query language}, especially due to  the large number of semantic problems they bring.

Second, even if a database would internally use several relations to represent a conceptual single table that would result from a join, it is not clear that this representation is computationally more expensive. This is precisely the underlying idea of \emph{factorized databases}~\cite{OlteanuZavodny2015}, which in fact can perform certain operations faster because they avoid materializing large results. Indeed, several products have now demonstrated that worst-case optimal join algorithms and factorization can temper this performance sacrifice. 

Third, if we do not have nulls in our query language, do we need to write  many joins in our queries? We can just define a view. For coding agents, simple design and consistency are more important.




\smallskip
\noindent \textbf{Debunk 3: We Need 3VL to Reason About Nulls.} Some believe that reasoning about incomplete information {\em requires} an expressive logic. At least that  was the reason SQL chose a 3-valued logic (we denote it here by 3VL, though there are many 3-valued logics out there, and SQL chose a particular one, known as Kleene’s logic). However, this is not the case. It is known that using 
the standard two-valued Boolean logic instead of 3VL makes no change in terms of expressiveness of SQL: whatever we could express with with 3VL, we can still express with Boolean logic. This was first shown for relational calculus queries \cite{CGL2022} and then extended to grouping, aggregation, and recursive queries \cite{pods23}.

To give an idea how  evaluation based on Boolean logic works, note that three different truth values arise in evaluating conditions in {\tt WHERE} and {\tt HAVING}. These appear typically in  comparisons conditions $e\, \omega\, e'$ where $\omega$ is one of $=, \neq, <, >, \leq, \geq$, and $e,e'$ are expressions producing values. Of these comparisons, three subsume equality: $=, \leq, \geq$. 
Now we can say that $e \ \omega\ e'$ is true if neither $e$  nor  $e'$ evaluates to \texttt{NULL} and  $e\ \omega\ e'$   holds in the usual sense, or if both $e$  and  $e'$ evaluate to  \texttt{NULL} and   $\omega$ subsumes equality. In all other cases (that is, when one argument is {\tt NULL} or both are and the comparisons are $\neq, <, >$), the result is false. Here we treat {\tt NULL} syntactically, the way grouping and set operations do in SQL. 

In fact it is not only this specific way of handling nulls with Boolean logic, but {\em every} way that is not unreasonable (e.g., saying that {\tt NULL} equals 42 and nothing else) gives us a language that is as expressive as SQL under 3VL \cite{pods23}. 


\subsection*{The Case for No Nulls and No 3VL}

We believe that query languages are better off without three-valued logic (3VL) and, more radically, without nulls altogether. 

\paragraph{Dispensing with 3VL}
Replacing 3VL with ordinary Boolean logic in query evaluation does not cost us anything in terms of expressiveness. At the same time, it brings several tangible benefits.

To start with, 
for the overwhelming majority of queries, the switch changes nothing at all:  3VL and Boolean evaluation agree on all TPC-H and TPC-DS queries but one---over 99\% of these standard benchmark queries. Their disagreement is rare. 

 Boolean logic removes a longstanding source of confusion around the third truth value \emph{unknown}. In SQL, its application depends on where it occurs: in the \texttt{WHERE}-clause and join conditions, \emph{unknown} is effectively treated as \emph{false} (only tuples evaluating to \emph{true} are kept), whereas in  constraints it is effectively treated as \emph{true} (a constraint holds if it is not  \emph{false}). A logic that needs two different conflations to make sense operationally is bound to create confusion leading to programming errors. 

Furthermore, Boolean logic restores useful algebraic identities that are lost under 3VL, but can facilitate query optimization. For instance, $\sigma_\theta(E) \cup \sigma_{\lnot\theta}(E) = E$ holds under Boolean logic but fails under 3VL, since tuples for which $\theta$ evaluates to \emph{unknown} are excluded from both sides. For a more detailed list of such recovered identities, related to {\tt ALL}, {\tt SOME}, and negation interactions, see \cite{pods23}. 
Last but not least, 
a user study showed that respondents prefer the Boolean semantics by
roughly a 2-to-1 margin \cite{pods23}, suggesting that few tears will be
shed if future query languages do not use 3VL.

\paragraph{Dispensing with nulls.}
Once 3VL is off the table, nulls lose much of their original motivation, and the more principled position is to do without nulls entirely.
 This is not a new proposal: it was already taken three decades ago by Date and Darwen in their Tutorial~D language~\cite{DarwenDate1995}, where missing information is instead represented explicitly, e.g. via default values or relation-valued attributes, rather than through a distinguished marker value threaded through every operator's semantics. We are now much better equipped to make this a reality. We have already shown that nulls can be eliminated as a  language concept, for example, by resorting to full normalization. The experience of the Rel language \cite{Rel} whose clean semantics simplifies reasoning tasks and reduces development costs while still delivering good performance, is a testament to the implementability of this principled approach to language design.

\OMIT{
\subsection*{The case for no nulls and no 3VL}

Start with no 3VL and move some stuff from above.

Then no nulls

Don't forget to cite tutorialD

\begin{itemize}
\item using Boolean logic instead of 3VL makes no change in terms of expressiveness of SQL: what we could say with 3VL, we can still say with Boolean logic;
\item 
For most queries, no change is necessary: whether we use 3VL or Boolean logic, the result is the same (this applies to all TPC-H and TPC-DS queries except one, i.e. over 99\%).
\item 
This also eliminates confusion many people have about the third value \emph{unknown} in queries and constraints: in evaluating conditions in queries it is in the end conflated with \emph{false} (so only true tuples are kept) but in conditions it is conflated with \emph{true} (all non-false tuples pass). 
\item 
On the other hand, Boolean logic allows us to recover some useful optimizations: for example, $\sigma_\theta(E) \cup \sigma_{\lnot\theta}(E) = E$.
\item 
Finally, a user survey shows that by roughly a 2-to-1 margin users prefer results given by evaluation based on Boolean logic. 
\end{itemize}
}

\section{Bag Semantics}
Codd's reference to bags as \emph{corrupted relations}~\cite{codd-book} is already sufficient to give us an impression of his opinion about them. So, why do our query languages have bag semantics?
Arguments for bag semantics typically fall into three categories:
\begin{enumerate}
\item We need bags for aggregation.
\item We need bags because multiplicities have semantics.
\item We need bags for performance (e.g., union).
\end{enumerate}
We will debunk arguments 1 and 2. While the third argument is true on the surface, we will argue that its disadvantages -- rarely discussed -- outweigh its benefits. 

We outline two further, fundamental issues of bag semantics. One is its interaction with iteration or recursion, turning simple computations into ones with astronomical (and beyond) outputs.
The other is the inability to provide a clear {\em declarative semantics} in the presence of bags: to explain how they are handled, we must rely on procedural languages. 

\smallskip
\noindent \textbf{Debunk 1: Bags Are Needed for Aggregation.}
This perceived need is due to thinking of aggregation as \emph{first project and then aggregate}. For example, {\tt SELECT SUM(B) FROM R} for a relation {\tt R(A,B)} with tuples (1,2), (2,2) may be thought of as:
\begin{quote} 
project on column \verb+B+ and create a \emph{bag} $\{\!\{2,2\}\!\}$\\
then apply the \verb+SUM+ aggregate to this bag.
\end{quote}
In principle, a simple change in SQL's formal semantics can solve this issue. 
Instead, this query can be defined as 
$$\sum_{t \in R}t.B$$
without ever producing a bag with projections on the $B$ column. In fact, query language design could stick to fundamental operators from the programming language community, and think about aggregation in terms of \verb+reduce+ or \verb+fold+ operations. Then, the above query is simply $\texttt{reduce}[0,f](R)$ which iterates over tuples in $R$ and adds $t.B$ to the accumulator value $acc$. Here $f(acc,t) = acc + t.B$. 
As a case in point, the language Rel~\cite{Rel} works entirely with set semantics and completely supports aggregation. 
For example, with relations {\tt DE} of departments and employees and {\tt ES} of employees and salaries, 
the average salaries in departments for which the sum of salaries are larger than 100 are computed by
\begin{verbatim}
def DES(d,e,s) : DE(d,e) and ES(e,s)
def q[d] : average[DES[d]] where sum[DES[d]] > 100    
\end{verbatim}

The application {\tt q[d]} effectively performs grouping; we refer the reader to \cite{Rel}. The key point is that the language never abandons set semantics: the join {\tt DES} is a set, and aggregations are computed as above, over entire sets, without projecting first.  Note that by itself this is not surprising: we have known for a long time that the power of relational languages  under bag semantics is the same as of languages under set semantics with added aggregation \cite{LibkinWong1997,GrumbachMilo1996}, even though it took several decades to turn this knowledge into a commercially viable implementation.

\smallskip
\noindent \textbf{Debunk 2: Multiplicities Should Have Semantics.}
This sometimes comes in more specific forms, such as needing bags for
\begin{enumerate}[(a)]
\item dealing with duplicate rows in the input data or
\item being able to output a list with duplicate rows.
\end{enumerate}
We first discuss these two cases before diving deeper. These 
arguments are specifically about \emph{input} and \emph{output}
behavior, which we should separate from the principles of our {\em
  data model}. Why? Because our many possible desired forms of query output should
not dictate what the data model is. Sometimes we may want database queries to
output an \emph{ordered list} (say, using  \verb+ORDER BY+). This does not mean that we should change the entire relational
database model to use ordered relations: the fact that relations are
unordered gives us the freedom to process and optimize queries how we
want. 
As argued in \cite{dontkillus}, a query language should never leave the data model to provide bespoke user representations for different tasks.
%
It is perfectly fine to separate input/output
behavior from the internal data model of systems.

There are more reasons why (a) and (b) should not convince us to use
bag semantics. If the source data has duplicate rows, it means that
some entity that should have been modeled is being overlooked. Thus, bag
semantics enables bad modeling. For example, if we read an input file
of person names that contains duplicates representing different
persons, we should generate person IDs that correspond to the {\em entities} (i.e., persons)
that these names belong to.  A similar argument
holds for (b). If we need to produce a list of names for some output,
but are bound to set semantics, we can simply produce unique
identifiers for the persons and output \verb+(ID+, \verb+name)+ 
pairs.

Finally, 
if multiplicities are important, it is 
somewhat primitive to represent them through multiple occurrences
of a tuple in a table, i.e., in space consuming unary encoding. Decimal or binary encodings seem more natural: instead of producing 42 copies of a tuple in the output,
produce the pair (name,42) or (name,101010). 

In any case, the reader should ask themselves if these points about consuming input and writing output outweigh the long-term advantages of set semantics, see the end of this section. 

\smallskip
\noindent \textbf{Debunk 3: Performance.}
A valid argument for bags is that they allow performance improvements for simple operations. In particular, inserting a tuple into a relation can be done without checking if it already exists. Taking the union of two relations can be done by simply appending tables. Projection can be done without duplicate elimination. In fact, this may be the exact reason why early 
systems used bags and why we still use them five decades later.

But are we amputating an arm for the sake of losing weight? While giving some performance gains on the surface, bag semantics costs us many opportunities for query optimization. Every optimization rewrite that is possible under bag semantics is also possible under set semantics. But the converse is not true. 
The classical example is optimization of {\em many-way joins}, or more generally, {conjunctive queries (CQs)}: 
\[
R_1 \bowtie R_2 \bowtie \cdots \bowtie R_n
\]
where $R_1, \dots, R_n$ are database relations. Under \emph{set semantics}, we know that  there is a unique optimal equivalent query \cite{ChandraMerlin1977}. It can be found in NP, based on NP-completeness of  containment $Q_1 \subseteq Q_2$ of CQs. 
 This result is very robust:
\begin{itemize}
    \item containment of unions of such queries, i.e.\ $CQ_1 \cup \cdots \cup CQ_m$, is also decidable in NP,
    \item so is containment when disequalities ($\neq$) can be used in join conditions (instead of just equijoins).
\end{itemize}
In the case of \emph{bags}, however:
\begin{itemize}
    \item for the general containment problem, its decidability is still {\em unknown} (and even if proved, it is likely the algorithm will be completely impractical),
    \item for the above extensions (union and disequalities), containment becomes \emph{undecidable},
    \item some cases remain decidable (e.g.\ no projection), 
    but algorithms are significantly more complex.
\end{itemize}

CQ containment
has been used in 
 a variety of tasks in data
integration \cite{DoanHalevyIves2012}, data exchange \cite{ArenasBarceloLibkinMurlak2014}, answering queries under ontologies \cite{BienvenuO15}, to name some. We lose these tools with
bags.

\medskip

\paragraph{Recursion.} With database tasks growing more diverse and complex, recursion
is more and more common. We now offer another thought
experiment, addressing the issues of both performance and the meaning of
multiplicities for simple recursive queries. 
As an example, we take the canonical recursive query computing the
{\em transitive closure} ({\it TC}) of a relation. There are two ways to write it:
we can connect the non-recursive and the recursive parts by
\verb+UNION+, forcing set semantics, or by \verb+UNION ALL+, forcing
bag semantics. 

\begin{figure}[t] 
  \centering
  \includegraphics[width=\linewidth]{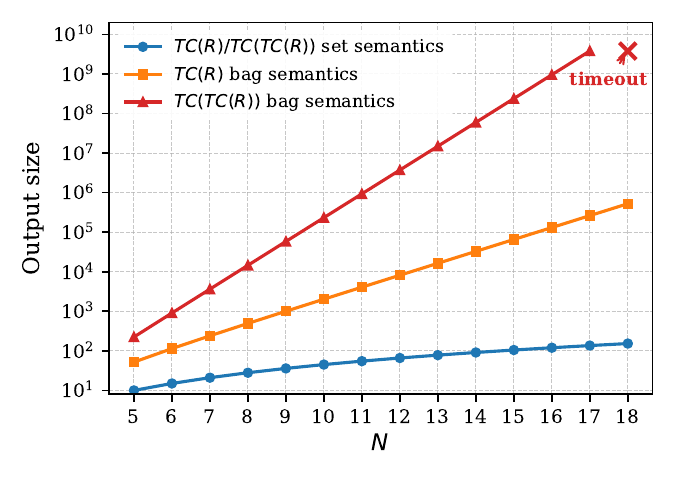}
  \vspace{-10mm}
  \caption{Growth of query outputs in set vs. bag semantics.}
  \label{fig:tc-growth}
\end{figure}

Assume that  $R$ contains tuples $(1,2), (2,3), \dots,
(N-1,N)$; to make it into a bag, we put two copies of each tuple into
$R$.   
We look at the transitive closure query, and also at a query that
applies the transitive closure {\em twice}, i.e.,
$\mathit{TC}(\mathit{TC}(R))$. Under set semantics, this changes
nothing: transitive closure of a relation is transitive, and hence 
$\mathit{TC}(\mathit{TC}(R)) = \mathit{TC}(R)$, with both having
$N(N-1)/2$ tuples. Under bags, however,
it is a completely different story. 
Figure~\ref{fig:tc-growth} plots, on the log scale, the sizes of 
$\mathit{TC}(R)$ and $\mathit{TC}(\mathit{TC}(R))$ under set and bag semantics. With set semantics and {\tt UNION}, we are under 200 tuples for $N=18$. With bag semantics, $\mathit{TC}(R)$ has  hundreds of thousands of tuples for the same $N$, and for $\mathit{TC}(\mathit{TC}(R))$ we are in the timeout land (with expected size roughly $1.5\cdot 10^{10})$; even for $N=17$ the size is well over a billion tuples. 
In fact, it takes only  $N=134$ (and thus 266 tuples in $R$) for the size of the output to exceed the number of protons in visible universe. 

\OMIT{
\paragraph{Transitive closure (TC)} We only return the count of tuples:

\begin{lstlisting}
WITH RECURSIVE tc(a, b) AS (
    SELECT * FROM r
    UNION
    SELECT r.a, tc.b
    FROM r, tc
    WHERE r.b = tc.a
)
SELECT COUNT(*) FROM tc;
\end{lstlisting}

\paragraph{TC of TC} Of course, mathematically this is just TC,  
since the transitive closure of a relation is transitive:

\begin{lstlisting}
WITH RECURSIVE tc(a, b) AS (
    SELECT * FROM r
    UNION
    SELECT r.a, tc.b
    FROM r, tc
    WHERE r.b = tc.a
),
tc_of_tc(a, b) AS (
    SELECT * FROM tc
    UNION
    SELECT tc.a, tc2.b
    FROM tc, tc_of_tc tc2
    WHERE tc.b = tc2.a
)
SELECT COUNT(*) FROM tc_of_tc;
\end{lstlisting}

Then we look at the \emph{bag version} of these,  
where each \texttt{UNION} is replaced by \texttt{UNION ALL}.  
In the set version, we always get $N \cdot (N-1)/2$ tuples in the output.  
But in the bag version we get exponential (and eventually astronomical) growth, see Figure~\ref{fig:tcoftc}.
}

\OMIT{
\begin{figure*}
\begin{tabular}{rrrrr}
\toprule
$N$ & size of TC & TC of TC sets & TC with bags & TC of TC with bags \\
\midrule
5  & 10  & 10  & 52     & 224 \\
6  & 15  & 15  & 114    & 906 \\
7  & 21  & 21  & 240    & 3,636 \\
8  & 28  & 28  & 494    & 14,558 \\
9  & 36  & 36  & 1,004  & 58,248 \\
10 & 45  & 45  & 2,026  & 233,010 \\
11 & 55  & 55  & 4,072  & 932,060 \\
12 & 66  & 66  & 8,166  & 3,728,262 \\
13 & 78  & 78  & 16,356 & 14,913,072 \\
14 & 91  & 91  & 32,738 & 59,652,314 \\
15 & 105 & 105 & 65,504 & 238,609,284 \\
16 & 120 & 120 & 131,038 & 954,437,166 \\
17 & 136 & 136 & 262,108 & 3,817,748,696 \\
18 & 153 & 153 & 524,250 & timeout (75min) \\
\bottomrule
\end{tabular}
\caption{Growth of transitive closure sizes in set vs.\ bag versions.}
\label{fig:tcoftc}
\end{figure*}
}

Even beyond computational issues, what is the significance of having 64 edges from 3 to 9 in the bag transitive closure or 2048 edges in the TC of TC? 
%
Playing devil's advocate, let us try to counter this argument by saying: no, these numbers have a meaning, they tell you how many times a tuple was derived. This looks reasonable on the surface but upon closer examination, this argument reveals another fundamental problem with bag semantics: it does not exist for {\em declarative languages}. The number of derivations of a tuple for a declarative query has no meaning per se until we translate it into a \emph{procedural} query, and such a translation is far from being unique. Diving deeper, note that early formal definitions of bag semantics  \cite{Albert1991,LibkinWong1997,GrumbachMilo1996} were given for {procedural} languages like relational algebra; even for datalog it was  defined procedurally, either in terms of counting proof trees or applying chase and reducing to sets  \cite{MumickPirahesh1990,BertossiGottlobPichler2019}. 

To understand why this is so, consider a logical query $q(x,z) = \exists y\ \big(R(x,y) \wedge S(y,z)\big)$, that we can also write as a Datalog rule $q(x,z) \mbox{:--} R(x,y),   S(y,z)$. The number of occurrences of a tuple $(a,b)$ in the output is given by $\sum_{c} \#_R(a,c)\cdot \#_S(c,b)$, where $\#_R(\bar u)$ denotes the multiplicity of a tuple $\bar u$ in relation $R$ \cite{HernichKolaitis2017}. Why this particular expression? Simply because the above conjunctive query written in logical notation translates into $R \Join_y S = \pi_{x,z}(R \times S)$, and the expression we used is the one that corresponds to the well-accepted bag semantics of 
Cartesian product and projection. In other words, our declarative bag semantics is driven by a particular translation from logic into algebra! But we know well that such a translation, in the set case, is by no means unique: even different textbooks present different logic-to-algebra translations. So we somehow end up depending on one a priori chosen translation, which looks as a bizarre choice, to say the least. Even if  for CQs it may pretend to have some element of ``canonicity'' (though this is questionable too), beyond CQs any choice of a translation into a procedural language  will have a huge element of arbitrariness in it, extending thus to the semantics of a declarative language itself. 


\subsection*{The Case for Sets}\label{sec:pro-sets}
In terms of query optimization, we already mentioned that every optimization rewrite that is possible for bags is also possible for sets. By sticking to bags, however, we are leaving many optimization opportunities, especially for join optimizations. 


Another point is consistency. As the scope of query languages continues to grow, it becomes increasingly difficult to apply bag semantics consistently. Recursive queries are becoming increasingly important in query languages for complex workloads involving graph analytics, ontological, predictive, and prescriptive reasoning. 
The perils of bag semantics are especially noticeable in graph languages such as SPARQL, and the newly released standards GQL and SQL/PGQ. In the former, bag semantics assigned to each path a multiplicity corresponding to how many times it was derived. This however resulted in simple queries with astronomical size output \cite{yottabyte}. The common pattern matching language of GQL and SQL/PGQ \cite{sigmod22}, well aware of this issue, imposed several restrictions. One of them limited the class of available regular expressions, in a way that reduced expressiveness of the language. This is a high price to pay for having duplicates. Worse yet, the pattern matching algorithm of the PGQ and GQL standards enforces duplicate elimination as one of its steps, to reduce the space of possible matches. However, it was not the last step, and subsequent steps, applying classical relational operators under bag semantics, reintroduce duplicates. Thus, sticking to bag semantics makes the number of duplicates essentially meaningless: it has no declarative semantics  --- it is a consequence of one specific procedural algorithm.





Occasionally,  one hears an argument ``if the user wants sets, they can
always use {\tt DISTINCT}''. This does not hold water however. Such an
approach 
leads to {\em mixing} sets and bags with
unpredictable semantics. Returning to the example from
Section~\ref{bad-bags:sec}, if we wrote {\tt SELECT DISTINCT}
instead of {\tt SELECT}, would it fix the double counting problem?
\OMIT{
\begin{lstlisting}
SELECT DISTINCT o.cust_id, SUM(o.amount) 
FROM orders o JOIN lineitem l ON o.id=l.o_id 
GROUP BY o.cust_id  
\end{lstlisting}
}
The answer is negative: {\tt DISTINCT} does not prevent
double-counting, 
as it happens in the join. By the time {\tt DISTINCT} is applied, the
aggregate is already computed, and its value is wrong. It is a much
more principled solution to 
use one semantics throughout, and given problems with bags that set
semantics avoids, it better be set semantics.   


\section{Concluding Remarks}
Chamberlin \cite{Chamberlin12} lists three main criticisms of SQL: nulls, duplicates (bags), and impedance
mismatch. It is almost like admitting that we actually know what is
wrong. In fact, \emph{why} does SQL have bags and nulls? Is this because of language design reasons or due to performance reasons? If the answer is performance only, is it a case of \emph{premature optimization being the root of all evil} \cite{Knuth74}?

Concerning bags, it seems that \emph{we only have them for performance reasons}: to efficiently compute unions and inserts. However, the performance costs of bags down the line are huge: bags tie our hands together when we want to do query optimization. Even worse, they are semantically problematic in the presence of recursion, which means that we are more likely to solve the wrong problem.

In our query languages, nulls and 3VL create way more problems than they solve. Going back to Boolean logic recovers optimizations and eliminates semantic inconsistencies. We now know that, in principle, a language based on the first principles --- relations as sets of tuples of atomic values --- can be designed, implemented, and successfully deployed. We should separate concerns: 
query languages should hide unnecessary complexity from the user, whether semantic or algorithmic. Nulls and bags may have their place in data structures and lower-level languages, but in query languages it is time to seriously consider getting rid of the old baggage that 
they are.


\bibliographystyle{abbrv}
\bibliography{main}

\end{document}